\documentclass[aps,pra,reprint,superscriptaddress,floatfix,showpacs]{revtex4-1}
\usepackage{blindtext}
\usepackage[utf8]{inputenc}
\usepackage[T1]{fontenc}
\usepackage{bm}
\usepackage{dcolumn}
\usepackage{graphicx}
\usepackage{subfigure}
\usepackage{xcolor}
\begin{document}

\title{Exploring the role of hyperbolicity in surface enhanced Raman sensing}

\author{Mihir Kumar Sahoo}
\author{Abhay Anand V S}
\author{Nihar Ranjan Sahoo}
\author{Anshuman Kumar}
\email{anshuman.kumar@iitb.ac.in}
\affiliation{Laboratory of Optics of Quantum Materials (LOQM), Department of Physics, IIT Bombay, Mumbai, 400076, Maharashtra, India}

\begin{abstract}
A plasmonic nanostructure-based substrate, serving as a surface-enhanced Raman scattering (SERS) substrate, enhances the Raman scattering of molecules. By employing an electron beam lithography followed by our recently developed nano-electroplating protocol, a gold nanorod array SERS substrate can be fabricated to detect lower molecular analyte concentrations, such as Rhodamine 6G (R6G) solution. As the critical dimensions of the nanorod array decrease, they exhibit hyperbolic metamaterial (HMM) characteristics with anisotropic permittivity behavior. In our study, we fabricated two sets of nanorod arrays: one in the HMM regime (140 nm periodicity) and the other in the non-HMM regime (400 nm periodicity), aiming to evaluate the performance of each set based on R6G detection. The obtained results are compared and analyzed using COMSOL simulation and Raman mapping and the role of hyperbolicity is discussed.

\end{abstract}

\maketitle


\section{Introduction}

A sensor senses a particular substance mainly based on its sensitivity, i.e., detectable change in input \cite{Ding2016}. A particular type of sensor using Raman spectroscopy gives various molecular information based on the inelastic scattering of incident laser energy due to the vibrational modes of the molecules. However, Raman scattering spectroscopy suffers from a small scattering cross-section compared to that of the fluorescence, e.g., for every 10\textsuperscript{6} or 10\textsuperscript{8} incident photons on a molecule, only one or less undergoes Raman scattering \cite{Smith2004}. There may not be a Raman signal for lower analyte concentrations. 

Plasmonic-based nanostructures can surpass the limit and enhance the scattering intensity (proportional to the fourth power of electric field) using the surface-enhanced Raman scattering (SERS) technique \cite{Campion1998}. The Raman signal enhancement factor may be defined as the product of excitation and emission enhancement factors \cite{Wang2017, Mueller2021, Son2022}. An optimal Raman-enhanced signal can be obtained by maximizing the excitation and emission enhancement factor, which requires nano-dimensional patterns; more specifically, nanoscale gaps or hotspots are interesting as the field concentrates in small mode volume \cite{Zhang2017, Chen2023}. Apart from dispersive propagating surface plasmon modes, a nondispersive localized surface plasmon resonance (LSPR), occurs due to nanoscale gaps or hotspots, which amplify the Raman signal effectively \cite{Mayer2011, Chen2023, Chan2012, Habib2020}. It has recently been reported that an enhancement in Raman scattering signal is possible using a hyperbolic metamaterial (HMM) structure \cite{https://doi.org/10.48550/arxiv.2306.01314, Liu2021, Zhao2021, Shafi2021, Wang2021, Wang2022, Liu2023}. 

A hyperbolic material \cite{Peragut2017, Liu2021, Lai2017, Shafi2021, StarkoBowes2015} behaves as a metal for specific direction of light propagation and acts as a dielectric for the other due to the negative and positive permittivity tensor components, giving extreme anisotropy. The dispersion relation in wavevector space for such a system forms a hyperboloid (as described in the supplementary file), hence the name hyperbolic material. The hyperbolic material supports propagating waves with arbitrary wave vectors, which enhances the local density of states \cite{Khan2022, Ahmadivand2021}. However, the hyperbolic material is usually unavailable naturally in the visible range; therefore, artificial design is mandatory for the HMM. The HMM may be designed by making parallel metallic nanorods in a dielectric medium (Type-I HMM) or by stacking alternate metallic and dielectric planner nano-layers (Type-II HMM)  \cite{Wang2022}. The resonances in Type-II HMMs are typically weaker than those in Type-I HMMs owing to the former being metallic in two directions, giving rise to high reflection of the incoming fields \cite{StarkoBowes2015}. Hence in this work, we focus on Type-I HMMs (without depositing gold nanoparticles on it), which to our knowledge, has never been employed as a SERS substrate before. In our previous work, we developed a unique recipe to obtain nanoscale patterns using EBL followed by nano-electroplating \cite{Sahoo2023} which has so far only been used for micro-scale patterns. Using this technique, we fabricate Type-I metamaterials in this work. 

To investigate the effect of hyperbolicity on the SERS substrate, we fabricated HMM- and a non-HMM-AuNR array without depositing gold nanoparticles on them. The critical dimensions of the AuNR array distinguish an HMM and non-HMM substrate. Rhodamine 6G (R6G) is a water-soluble organic laser dye that exhibits an excellent spectroscopic property typically employed for characterizing SERS substrates. We tune the filling factor (FF) of the metamaterial to explore the influence of hyperbolicity at the excitation wavelength. We perform experiments with larger periodicity arrays varying their FF. Our results indicate that while there is SERS enhancement in both cases compared to planar substrates, the enhancement in going from HMM to non-HMM for all FFs is comparable, which indicates that hyperbolicity may not provide extraordinary SERS enhancement compared to LSPR. Since for the visible range, the required feature size of HMM is extremely small, it may not be helpful to go for these HMMs for the purposes of SERS enhancement.

\begin{figure*}
    \centering
    \includegraphics[width = 1\textwidth]{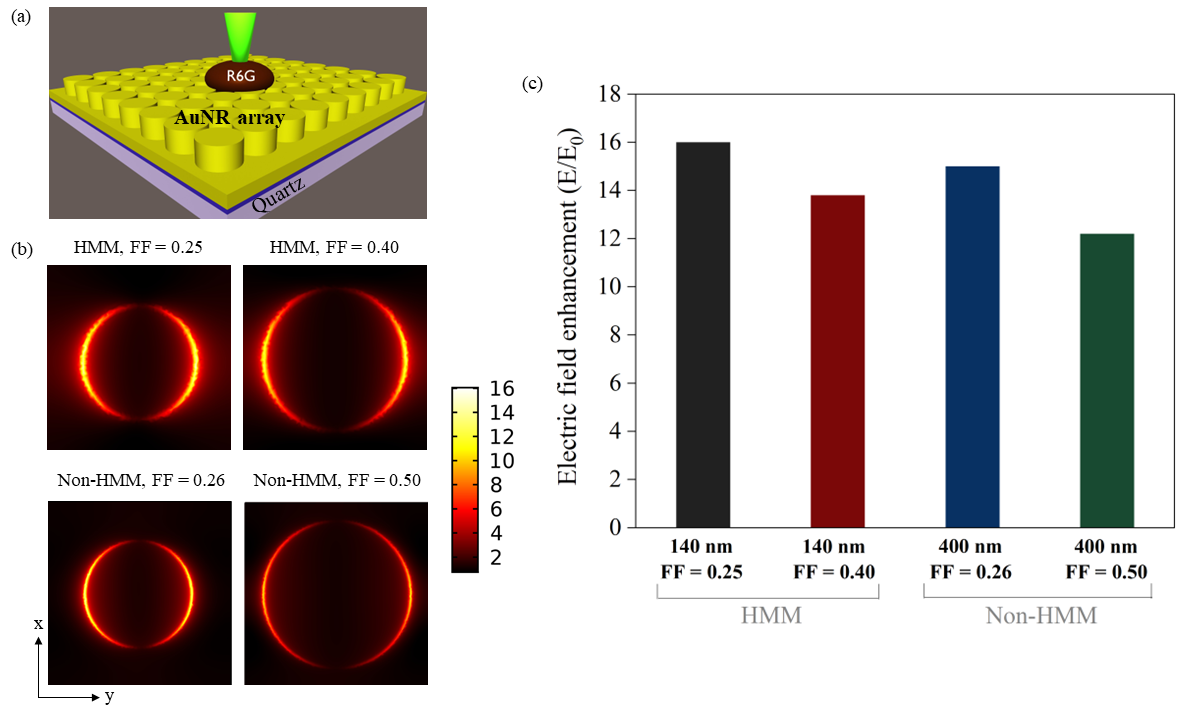}
    \caption{(a) Schematic of AuNR array fabricated on quartz substrate via electron beam lithography (EBL) followed by Au electroplating protocol. The electroplating requires a seed layer (Au film), and a Ti-adhesive layer holds the Au film on the cleaned quartz substrate. (b) Simulated HMM- and non-HMM-AuNR array for an incident wavelength of 532 nm. (c) Comparison of simulated electric field enhancement for various FFs of AuNR array.}
    \label{FIG:1}
\end{figure*}

\section{Methods}
Fig.~\ref{FIG:1} (a) shows the schematic of the AuNR array fabricated  using EBL and electroplating protocol. A detailed procedure for fabricating the AuNR array was given in our previous work \cite{Sahoo2023}. The height of the AuNR array is 65 nm, as obtained from the AFM analysis (described in supplementary section 1, Figure S1). For 532 nm excitation, a 140 nm periodicity with various FFs is considered as an HMM-AuNR array, and a non-HMM AuNR array is obtained from a 400 nm periodicity with various FFs. Note that for our excitation wavelength, the HMM periodicity and the corresponding height are consistent with the practical validity range of the effective medium approximation as reported elsewhere \cite{Wang2022, Vasilantonakis2015}. The FF can be calculated from the radius (r) and periodicity of the AuNR array as FF = $\pi(r/p)\textsuperscript{2}$. Fig.~\ref{FIG:1} (b) shows COMSOL simulation results of electric filed distribution on HMM- and non-HMM-AuNR array in the x-y plane for an incident wavelength of 532 nm. Fig.~\ref{FIG:1} (c) shows the comparison of simulated electric field enhancement for various FF of the AuNR array. Supplementary section 2 (Figure S2) describes the simulation results of the AuNR array along the vertical direction (z-axis) for a height of 65 nm.  

The R6G powder (C\textsubscript{28}H\textsubscript{30}N\textsubscript{2}O\textsubscript{3}HCl, dye content nearly 99\%, product number-R4127, molecular weight = 479.01 g per mol) was purchased from Sigma-Aldrich. A 4.79 g of R6G powder was put in 10 ml DI water to prepare one mM R6G solution. Ten $\mu$l of one mM solution was poured in 9.990 ml of DI water to prepare one $\mu$M R6G solution using serial-wise mixing (C\textsubscript{1}V\textsubscript{1} = C\textsubscript{2}V\textsubscript{2}). The R6G solutions (one $\mu$M) were prepared and drop-casted on the fabricated AuNR array (as the SERS substrate) to obtain an enhanced Raman signal.

\begin{figure*}
    \centering
    \includegraphics[width= 1\textwidth]{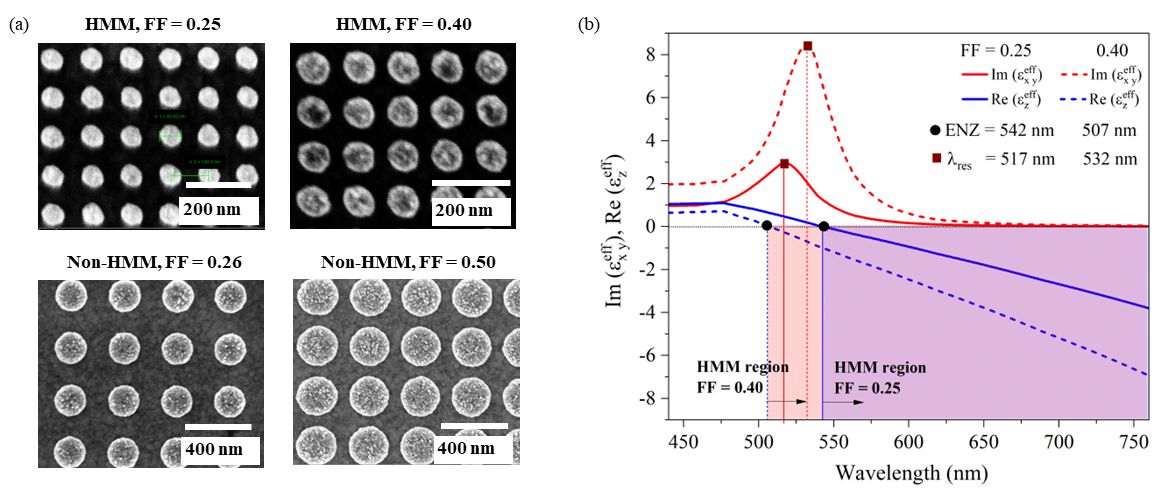}
    \caption{(a) SEM images of HMM and non-HMM AuNR array fabricated for various FFs. (b) Effective permittivity ($\epsilon_{x, y}^{eff}$, $\epsilon_z^{eff}$) are calculated analytically using effective medium theory (EMT) in R6G medium (n = 1.33) for all the FFs of HMM structure only (as the EMT is not valid for non-HMM structures). The epsilon near zero (ENZ) region is the zero-crossing of Re ($\epsilon_z^{eff}$) curve, after which HMM region proceeds and the resonance wavelength ($\lambda_{res}$) occurs at the peak of Im ($\epsilon_{x, y}^{eff}$) curve. For FF = 0.25, ENZ and $\lambda_{res}$ occur at 542 nm and 517 nm wavelength; and for FF = 0.40, ENZ and $\lambda_{res}$ occur at 507 nm and 532 nm wavelength, respectively. Solid- and dash-line indicate 0.25 and 0.40 FF of HMM AuNR array.}
    \label{FIG:2}
\end{figure*}

\begin{table*}[]
\renewcommand{\arraystretch}{1.8}
\caption {Measured parameters (periodicity (p) and radius (r)) of AuNR array and their calculated filling factors (FFs).}

\begin{tabular}{|c|c|c|}
\hline
\textbf{Periodicity (nm)} & \textbf{Radius (nm)} & \textbf{Filling factor (FF)} \\ \hline
140                       & 40                   & 0.25                         \\ \hline
140                       & 50                   & 0.40                         \\ \hline
400                       & 150                  & 0.44                         \\ \hline
400                       & 160                  & 0.50                         \\ \hline
\end{tabular}
\end{table*}

\section{Characterization}
SEM (make: Raith GmbH/150-Two), AFM (make: Bruker, Multimode Nanoscope-IV), XPS (make: ULVAC-PHI/PHI5000VersaProbeII), Filmetrics Reflectometer (F40 UV-VIS), and Raman (make: HORIBA/LabRAM HR Evolution)  characterizations were performed to measure diameter and spacing, height, verify Au deposition, refractive indices of Au thin film, and acquire Raman spectra of AuNR array fabricated using EBL and electroplating protocol. The Raman mappings were acquired through a micro-spectrometer equipped with an Ar ion laser ($\lambda$ = 532 nm) and a liquid nitrogen-cooled coupled charge device detector for 4 s acquisition time for all the measurements. The spectra resolution was set at 0.5 cm\textsuperscript{-1} (1800 grooves per mm grating), and the laser was focused on the sample using 100X objective with 1.4 mW laser power.

\begin{figure*}
    \centering
    \includegraphics[width=1\textwidth]{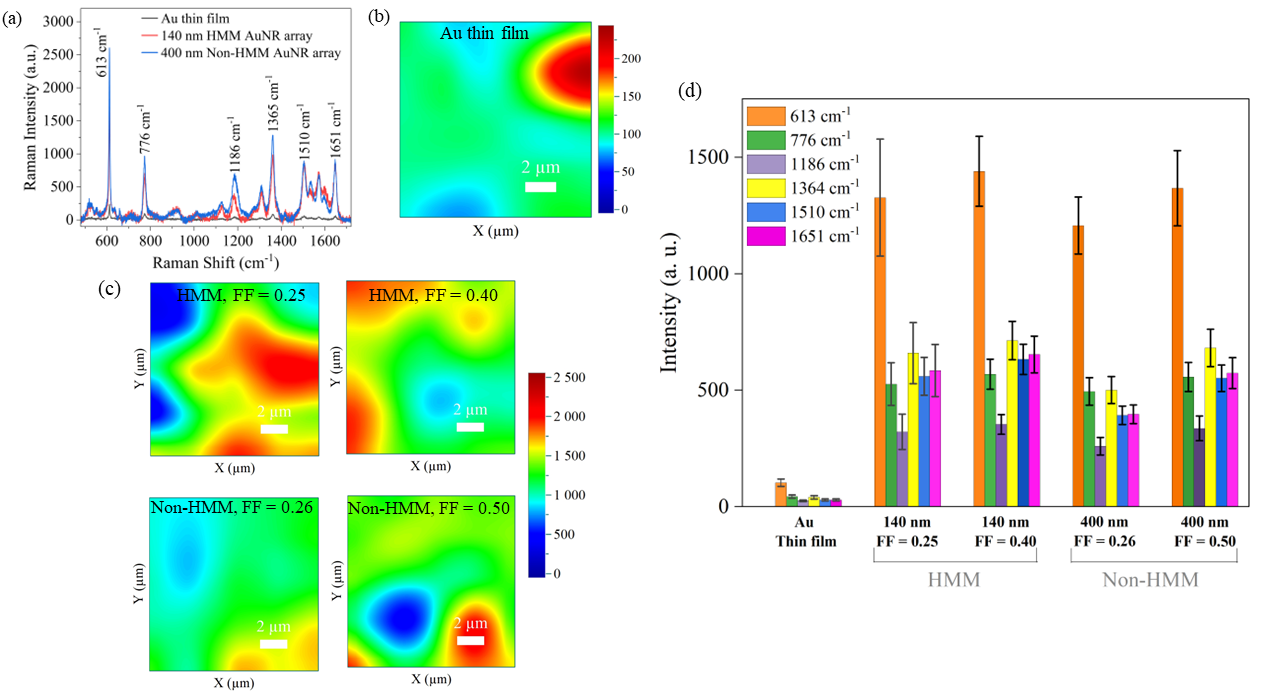}
    \caption{(a) Raman spectra of R6G molecules (one $\mu$M) detected from Au thin film, HMM-, and Non-HMM-AuNR array. Raman mapping of R6G molecules (one $\mu$M) was carried, and the highest intensity R6G peak (i.e., 613 cm\textsuperscript{-1}) was plotted  for (b) Au thin film, and (c) various FFs of HMM-, and Non-HMM-AuNR array. (d) Raman intensity of R6G at 613 cm\textsuperscript{-1}, 776 cm\textsuperscript{-1}, 1186 cm\textsuperscript{-1}, 1364 cm\textsuperscript{-1}, 1510 cm\textsuperscript{-1}, and 1651 cm\textsuperscript{-1} peaks detected from Au thin film, and various FFs of HMM-, and Non-HMM-AuNR array.}
    \label{FIG:3}
\end{figure*}

\section{Results}
Fig.~\ref{FIG:1} (b) and Fig.~\ref{FIG:1} (c) show simulation results of distribution and enhancement of electric field in the AuNR array. The simulation results show high electric field enhancement in both cases. However, the enhancement factor of the electric field in the HMM-AuNR array is comparable to the non-HMM-AuNR array.
Fig.~\ref{FIG:2} (a) shows the surface view of SEM images of a uniformly patterned HMM- and non-HMM-AuNR array fabricated using EBL and electroplating protocols for FFs. XPS measurements were performed on the electroplated Au thin film and AuNR array to verify the Au deposition (as described in the supplementary section 3, Figure S3) in electroplating technique. Table 1 shows measured AuNR array physical parameters from which the FFs are calculated. Fig.~\ref{FIG:2} (b) shows effective permittivity ($\epsilon_{x, y}^{eff}$, $\epsilon_z^{eff}$) of the HMM-AuNR array in the R6G host medium (in the visible region). A detailed explanation of effective permittivity calculation using analytical techniques is described in supplementary section 4 (Figure S4 and S5). 

One $\mu$M R6G solution was drop-casted on the samples to obtain the SERS signal from the Au thin film and the AuNR array at various FFs of HMM and non-HMM structure, as shown in Fig.\ref{FIG:3}(a). The Raman spectra highlight all the vibrational modes of R6G molecules (as described in the supplementary section 5). Compared to the Au thin film, the AuNR array exhibits a stronger Raman signal enhancement highlighting the SERS substrate in detecting lower molecular concentrations. To distinguish the SERS enhancement, Raman spectra were taken at two similar FFs of the HMM- and non-HMM-AuNR array, which results comparable enhancement. Instead of Raman spectra at a particular point, Raman mapping was carried out to obtain the SERS enhancement in an area (as shown in optical images of supplementary section 5, Figure S6) on the AuNR array. Among all the vibrational modes of the R6G molecule, the 613 cm\textsuperscript{-1} peak gives the highest enhancement and hence is mentioned here for Raman mapping. Fig. \ref{FIG:3} (b) shows Raman mapping of R6G molecule on the Au thin film for 613 cm\textsuperscript{-1} peak. Fig. \ref{FIG:3} (c) shows Raman mapping of R6G molecule on various FFs of HMM-and non-HMM-AuNR array for 613 cm\textsuperscript{-1} peak. Fig. \ref{FIG:3} (d) shows a Raman intensity contrast graph for Au thin film and various FFs of HMM- and non-HMM-AuNR array.

\section{Discussion}
 To investigate the crucial role of HMM property on SERS enhancement, a non-HMM-AuNR array (400 nm periodicity) was considered, and compared with the SERS performance with the hyperbolic one. The COMSOL simulation results in Fig.\ref{FIG:1} (b) highlight the enhanced and localized electric field distribution for various FFs of HMM- and non-HMM-AuNR array, exhibiting a comparable SERS enhancement (as shown in Fig.\ref{FIG:1} (c)). Supplementary section 2 shows the simulated field profile in the vertical direction, where an enhanced localized electric field occurs at the two corners of the AuNR, as observed in the surface view (Fig.\ref{FIG:1} (b)). However, in both sets of AuNR array, a hybridization of localized and surface plasmon polariton takes place at the base of the AuNR array (as shown in supplementary Figure S2) that also exhibits a comparable electric field enhancement in HMM- and non-HMM-AuNR array.        

To investigate the dependency of Raman enhancement on anisotropy and hyperbolicity, the Raman mapping of the R6G molecule was taken from two FFs of the AuNR array \ref{FIG:3} (b). In Type-I HMM, the crossing of Re ($\epsilon_{z}^{eff}$) curve with the epsilon-zero line gives the epsilon near zero (ENZ) region beyond which the HMM region appears, and the peak of the Im (($\epsilon_{x, y}^{eff}$) curve provides the resonance wavelength $\lambda$\textsubscript{res} \cite{StarkoBowes2015}, as shown in \ref{FIG:2} (b). For lower FF (0.25), $\lambda$\textsubscript{res} is away from the HMM region; therefore, the Raman intensity is lower compared to 0.40 FF, as shown in Fig.\ref{FIG:3} (b). However, the Raman intensity enhances as the $\lambda$\textsubscript{res} matches with the excitation wavelength at 532 nm and also falls in HMM region for 0.40 FF.  

We tried to fabricate similar FFs of the AuNR array using EBL and electroplating for 400 nm periodicity in order to compare the SERS performances with the 140 nm periodicity. Out of two FFs, one FF is nearly comparable, as mentioned in Table 1. The EMT approximation may fail at higher dimensions of the AuNR array, i.e., for 532 nm excitation, a 400 nm periodicity may not exhibit subwavelength criteria, therefore, we named it a non-HMM-AuNR array. The effective permittivity cannot be calculated using EMT for the non-HMM-AuNR array for 532 nm excitation, therefore, the Raman mapping results were compared for equivalent FFs of HMM- and non-HMM-AuNR array, e.g., 0.25 and 0.40 FFs of HMM-AuNR array, which may be compared with 0.26 and 0.50 FFs of the non-HMM-AuNR array, respectively. It must be noted that among the HMM-AuNR arrays, only FF of 0.40 exhibits hyperbolicity at our excitation wavelength. 

We first compare the elliptical region of the HMM-AuNR array with a similar FF of the non-HMM-AuNR array: the 0.25 FF of HMM exhibits a few times higher Raman intensity compared to the 0.26 FF of non-HMM. Similarly, while comparing the Raman intensity of 0.40 FF of HMM- (which is hyperbolic at our excitation wavelength) and 0.50 FF of non-HMM-AuNR array, the results show comparable Raman enhancement, which can be seen from the Raman mapping as well as a histogram comparison graph (Fig. \ref{FIG:3}). Other recent reports \cite{Liu2021, Shafi2021, Wang2021} also show a decrease or inadequate electric field enhancement in HMM compared to non-HMM (or metal nanostructures), which raises a question regarding the role of hyperbolicity in the SERS performance. It must be noted that to enhance the Raman signal, gold nanoparticles are deposited on the HMM sample, which again highlights the role of LSPR, not the HMM property, in enhancing the sensitivity of the SERS substrate. 

\textcolor{black}{It should also be noted that the strength of the Raman signal also depends on how the analyte spreads on the SERS substrate, so the observation is based on the average signal. It is possible that the observed trend is convoluted due to two simultaneously acting mechanisms of the structural profile of the analyte on the SERS substrate and the electric field enhancement effect, which requires further study.}

\section{Conclusion}
A low-cost and reproducible Au deposition via nano-electroplating forms a uniform AuNR array that is used as a SERS substrate. Detection of lower-concentration molecules, e.g., one $\mu$M R6G, via Raman spectroscopy, becomes easy on the AuNR array SERS substrate compared to the Au thin film. Tuning the FF of the AuNR array enhances the electric field and hence the Raman intensity of the R6G molecule. Comparison of average Raman intensity of R6G molecule between the HMM- and non-HMM-AuNR array exhibits inadequate Raman enhancement in Type-I HMM structure. The LSPR occurs due to the nanogaps/hotspots (among the nanorods) may only help in enhancing the SERS performance. The study raises a question regarding the candidature of pristine HMM (without depositing gold nanoparticles) in SERS functioning.    
\bibliography{sample}

\begin{thebibliography}{25}%
\makeatletter
\providecommand \@ifxundefined [1]{%
 \@ifx{#1\undefined}
}%
\providecommand \@ifnum [1]{%
 \ifnum #1\expandafter \@firstoftwo
 \else \expandafter \@secondoftwo
 \fi
}%
\providecommand \@ifx [1]{%
 \ifx #1\expandafter \@firstoftwo
 \else \expandafter \@secondoftwo
 \fi
}%
\providecommand \natexlab [1]{#1}%
\providecommand \enquote  [1]{``#1''}%
\providecommand \bibnamefont  [1]{#1}%
\providecommand \bibfnamefont [1]{#1}%
\providecommand \citenamefont [1]{#1}%
\providecommand \href@noop [0]{\@secondoftwo}%
\providecommand \href [0]{\begingroup \@sanitize@url \@href}%
\providecommand \@href[1]{\@@startlink{#1}\@@href}%
\providecommand \@@href[1]{\endgroup#1\@@endlink}%
\providecommand \@sanitize@url [0]{\catcode `\\12\catcode `\$12\catcode
  `\&12\catcode `\#12\catcode `\^12\catcode `\_12\catcode `\%12\relax}%
\providecommand \@@startlink[1]{}%
\providecommand \@@endlink[0]{}%
\providecommand \url  [0]{\begingroup\@sanitize@url \@url }%
\providecommand \@url [1]{\endgroup\@href {#1}{\urlprefix }}%
\providecommand \urlprefix  [0]{URL }%
\providecommand \Eprint [0]{\href }%
\providecommand \doibase [0]{http://dx.doi.org/}%
\providecommand \selectlanguage [0]{\@gobble}%
\providecommand \bibinfo  [0]{\@secondoftwo}%
\providecommand \bibfield  [0]{\@secondoftwo}%
\providecommand \translation [1]{[#1]}%
\providecommand \BibitemOpen [0]{}%
\providecommand \bibitemStop [0]{}%
\providecommand \bibitemNoStop [0]{.\EOS\space}%
\providecommand \EOS [0]{\spacefactor3000\relax}%
\providecommand \BibitemShut  [1]{\csname bibitem#1\endcsname}%
\let\auto@bib@innerbib\@empty
\bibitem [{\citenamefont {Ding}\ \emph {et~al.}(2016)\citenamefont {Ding},
  \citenamefont {Yi}, \citenamefont {Li}, \citenamefont {Ren}, \citenamefont
  {Wu}, \citenamefont {Panneerselvam},\ and\ \citenamefont {Tian}}]{Ding2016}%
  \BibitemOpen
  \bibfield  {author} {\bibinfo {author} {\bibfnamefont {S.-Y.}\ \bibnamefont
  {Ding}}, \bibinfo {author} {\bibfnamefont {J.}~\bibnamefont {Yi}}, \bibinfo
  {author} {\bibfnamefont {J.-F.}\ \bibnamefont {Li}}, \bibinfo {author}
  {\bibfnamefont {B.}~\bibnamefont {Ren}}, \bibinfo {author} {\bibfnamefont
  {D.-Y.}\ \bibnamefont {Wu}}, \bibinfo {author} {\bibfnamefont
  {R.}~\bibnamefont {Panneerselvam}}, \ and\ \bibinfo {author} {\bibfnamefont
  {Z.-Q.}\ \bibnamefont {Tian}},\ }\href {\doibase 10.1038/natrevmats.2016.21}
  {\bibfield  {journal} {\bibinfo  {journal} {Nature Reviews Materials}\
  }\textbf {\bibinfo {volume} {1}} (\bibinfo {year} {2016}),\
  10.1038/natrevmats.2016.21}\BibitemShut {NoStop}%
\bibitem [{\citenamefont {Smith}\ and\ \citenamefont {Dent}(2004)}]{Smith2004}%
  \BibitemOpen
  \bibfield  {author} {\bibinfo {author} {\bibfnamefont {E.}~\bibnamefont
  {Smith}}\ and\ \bibinfo {author} {\bibfnamefont {G.}~\bibnamefont {Dent}},\
  }\href {\doibase 10.1002/0470011831} {\emph {\bibinfo {title} {Modern Raman
  Spectroscopy {\textendash} A Practical Approach}}}\ (\bibinfo  {publisher}
  {Wiley},\ \bibinfo {year} {2004})\BibitemShut {NoStop}%
\bibitem [{\citenamefont {Campion}\ and\ \citenamefont
  {Kambhampati}(1998)}]{Campion1998}%
  \BibitemOpen
  \bibfield  {author} {\bibinfo {author} {\bibfnamefont {A.}~\bibnamefont
  {Campion}}\ and\ \bibinfo {author} {\bibfnamefont {P.}~\bibnamefont
  {Kambhampati}},\ }\href {\doibase 10.1039/a827241z} {\bibfield  {journal}
  {\bibinfo  {journal} {Chemical Society Reviews}\ }\textbf {\bibinfo {volume}
  {27}},\ \bibinfo {pages} {241} (\bibinfo {year} {1998})}\BibitemShut
  {NoStop}%
\bibitem [{\citenamefont {Wang}\ \emph {et~al.}(2017)\citenamefont {Wang},
  \citenamefont {Zhou},\ and\ \citenamefont {Li}}]{Wang2017}%
  \BibitemOpen
  \bibfield  {author} {\bibinfo {author} {\bibfnamefont {Y.}~\bibnamefont
  {Wang}}, \bibinfo {author} {\bibfnamefont {J.}~\bibnamefont {Zhou}}, \ and\
  \bibinfo {author} {\bibfnamefont {J.}~\bibnamefont {Li}},\ }\href {\doibase
  10.1002/smtd.201700197} {\bibfield  {journal} {\bibinfo  {journal} {Small
  Methods}\ }\textbf {\bibinfo {volume} {1}},\ \bibinfo {pages} {1700197}
  (\bibinfo {year} {2017})}\BibitemShut {NoStop}%
\bibitem [{\citenamefont {Mueller}\ \emph {et~al.}(2021)\citenamefont
  {Mueller}, \citenamefont {Pfitzner}, \citenamefont {Okamura}, \citenamefont
  {Gordeev}, \citenamefont {Kusch}, \citenamefont {Lange}, \citenamefont
  {Heberle}, \citenamefont {Schulz},\ and\ \citenamefont
  {Reich}}]{Mueller2021}%
  \BibitemOpen
  \bibfield  {author} {\bibinfo {author} {\bibfnamefont {N.~S.}\ \bibnamefont
  {Mueller}}, \bibinfo {author} {\bibfnamefont {E.}~\bibnamefont {Pfitzner}},
  \bibinfo {author} {\bibfnamefont {Y.}~\bibnamefont {Okamura}}, \bibinfo
  {author} {\bibfnamefont {G.}~\bibnamefont {Gordeev}}, \bibinfo {author}
  {\bibfnamefont {P.}~\bibnamefont {Kusch}}, \bibinfo {author} {\bibfnamefont
  {H.}~\bibnamefont {Lange}}, \bibinfo {author} {\bibfnamefont
  {J.}~\bibnamefont {Heberle}}, \bibinfo {author} {\bibfnamefont
  {F.}~\bibnamefont {Schulz}}, \ and\ \bibinfo {author} {\bibfnamefont
  {S.}~\bibnamefont {Reich}},\ }\href {\doibase 10.1021/acsnano.1c00352}
  {\bibfield  {journal} {\bibinfo  {journal} {{ACS} Nano}\ }\textbf {\bibinfo
  {volume} {15}},\ \bibinfo {pages} {5523} (\bibinfo {year}
  {2021})}\BibitemShut {NoStop}%
\bibitem [{\citenamefont {Son}\ \emph {et~al.}(2022)\citenamefont {Son},
  \citenamefont {Kim}, \citenamefont {Lee}, \citenamefont {Lee}, \citenamefont
  {Cha},\ and\ \citenamefont {Nam}}]{Son2022}%
  \BibitemOpen
  \bibfield  {author} {\bibinfo {author} {\bibfnamefont {J.}~\bibnamefont
  {Son}}, \bibinfo {author} {\bibfnamefont {G.-H.}\ \bibnamefont {Kim}},
  \bibinfo {author} {\bibfnamefont {Y.}~\bibnamefont {Lee}}, \bibinfo {author}
  {\bibfnamefont {C.}~\bibnamefont {Lee}}, \bibinfo {author} {\bibfnamefont
  {S.}~\bibnamefont {Cha}}, \ and\ \bibinfo {author} {\bibfnamefont {J.-M.}\
  \bibnamefont {Nam}},\ }\href {\doibase 10.1021/jacs.2c05950} {\bibfield
  {journal} {\bibinfo  {journal} {Journal of the American Chemical Society}\
  }\textbf {\bibinfo {volume} {144}},\ \bibinfo {pages} {22337} (\bibinfo
  {year} {2022})}\BibitemShut {NoStop}%
\bibitem [{\citenamefont {Zhang}\ \emph {et~al.}(2017)\citenamefont {Zhang},
  \citenamefont {Shen}, \citenamefont {Xie}, \citenamefont {Dou}, \citenamefont
  {Min}, \citenamefont {Lei}, \citenamefont {Liu}, \citenamefont {Zhu},\ and\
  \citenamefont {Yuan}}]{Zhang2017}%
  \BibitemOpen
  \bibfield  {author} {\bibinfo {author} {\bibfnamefont {Y.}~\bibnamefont
  {Zhang}}, \bibinfo {author} {\bibfnamefont {J.}~\bibnamefont {Shen}},
  \bibinfo {author} {\bibfnamefont {Z.}~\bibnamefont {Xie}}, \bibinfo {author}
  {\bibfnamefont {X.}~\bibnamefont {Dou}}, \bibinfo {author} {\bibfnamefont
  {C.}~\bibnamefont {Min}}, \bibinfo {author} {\bibfnamefont {T.}~\bibnamefont
  {Lei}}, \bibinfo {author} {\bibfnamefont {J.}~\bibnamefont {Liu}}, \bibinfo
  {author} {\bibfnamefont {S.}~\bibnamefont {Zhu}}, \ and\ \bibinfo {author}
  {\bibfnamefont {X.}~\bibnamefont {Yuan}},\ }\href {\doibase
  10.1039/c7nr02406a} {\bibfield  {journal} {\bibinfo  {journal} {Nanoscale}\
  }\textbf {\bibinfo {volume} {9}},\ \bibinfo {pages} {10694} (\bibinfo {year}
  {2017})}\BibitemShut {NoStop}%
\bibitem [{\citenamefont {Chen}\ \emph {et~al.}(2023)\citenamefont {Chen},
  \citenamefont {Zhang}, \citenamefont {Tang}, \citenamefont {Ye},
  \citenamefont {Zhang}, \citenamefont {Wu}, \citenamefont {Wang},
  \citenamefont {Zhang},\ and\ \citenamefont {Yang}}]{Chen2023}%
  \BibitemOpen
  \bibfield  {author} {\bibinfo {author} {\bibfnamefont {J.}~\bibnamefont
  {Chen}}, \bibinfo {author} {\bibfnamefont {C.}~\bibnamefont {Zhang}},
  \bibinfo {author} {\bibfnamefont {F.}~\bibnamefont {Tang}}, \bibinfo {author}
  {\bibfnamefont {X.}~\bibnamefont {Ye}}, \bibinfo {author} {\bibfnamefont
  {Y.}~\bibnamefont {Zhang}}, \bibinfo {author} {\bibfnamefont
  {J.}~\bibnamefont {Wu}}, \bibinfo {author} {\bibfnamefont {K.}~\bibnamefont
  {Wang}}, \bibinfo {author} {\bibfnamefont {N.}~\bibnamefont {Zhang}}, \ and\
  \bibinfo {author} {\bibfnamefont {L.}~\bibnamefont {Yang}},\ }\href {\doibase
  10.3390/coatings13050844} {\bibfield  {journal} {\bibinfo  {journal}
  {Coatings}\ }\textbf {\bibinfo {volume} {13}},\ \bibinfo {pages} {844}
  (\bibinfo {year} {2023})}\BibitemShut {NoStop}%
\bibitem [{\citenamefont {Mayer}\ and\ \citenamefont
  {Hafner}(2011)}]{Mayer2011}%
  \BibitemOpen
  \bibfield  {author} {\bibinfo {author} {\bibfnamefont {K.~M.}\ \bibnamefont
  {Mayer}}\ and\ \bibinfo {author} {\bibfnamefont {J.~H.}\ \bibnamefont
  {Hafner}},\ }\href {\doibase 10.1021/cr100313v} {\bibfield  {journal}
  {\bibinfo  {journal} {Chemical Reviews}\ }\textbf {\bibinfo {volume} {111}},\
  \bibinfo {pages} {3828} (\bibinfo {year} {2011})}\BibitemShut {NoStop}%
\bibitem [{\citenamefont {Chan}\ \emph {et~al.}(2012)\citenamefont {Chan},
  \citenamefont {Li}, \citenamefont {Ong}, \citenamefont {Xu},\ and\
  \citenamefont {Waye}}]{Chan2012}%
  \BibitemOpen
  \bibfield  {author} {\bibinfo {author} {\bibfnamefont {C.~Y.}\ \bibnamefont
  {Chan}}, \bibinfo {author} {\bibfnamefont {J.}~\bibnamefont {Li}}, \bibinfo
  {author} {\bibfnamefont {H.~C.}\ \bibnamefont {Ong}}, \bibinfo {author}
  {\bibfnamefont {J.~B.}\ \bibnamefont {Xu}}, \ and\ \bibinfo {author}
  {\bibfnamefont {M.~M.~Y.}\ \bibnamefont {Waye}},\ }in\ \href {\doibase
  10.1007/978-3-642-20620-7_1} {\emph {\bibinfo {booktitle} {Raman Spectroscopy
  for Nanomaterials Characterization}}}\ (\bibinfo  {publisher} {Springer
  Berlin Heidelberg},\ \bibinfo {year} {2012})\ pp.\ \bibinfo {pages}
  {1--32}\BibitemShut {NoStop}%
\bibitem [{\citenamefont {Habib}\ \emph {et~al.}(2020)\citenamefont {Habib},
  \citenamefont {Briukhanova}, \citenamefont {Das}, \citenamefont {Yildiz},\
  and\ \citenamefont {Caglayan}}]{Habib2020}%
  \BibitemOpen
  \bibfield  {author} {\bibinfo {author} {\bibfnamefont {M.}~\bibnamefont
  {Habib}}, \bibinfo {author} {\bibfnamefont {D.}~\bibnamefont {Briukhanova}},
  \bibinfo {author} {\bibfnamefont {N.}~\bibnamefont {Das}}, \bibinfo {author}
  {\bibfnamefont {B.~C.}\ \bibnamefont {Yildiz}}, \ and\ \bibinfo {author}
  {\bibfnamefont {H.}~\bibnamefont {Caglayan}},\ }\href {\doibase
  10.1515/nanoph-2020-0245} {\bibfield  {journal} {\bibinfo  {journal}
  {Nanophotonics}\ }\textbf {\bibinfo {volume} {9}},\ \bibinfo {pages} {3637}
  (\bibinfo {year} {2020})}\BibitemShut {NoStop}%
\bibitem [{\citenamefont {Lobet}\ \emph {et~al.}(2023)\citenamefont {Lobet},
  \citenamefont {Kinsey}, \citenamefont {Liberal}, \citenamefont {Caglayan},
  \citenamefont {Arroyo-Huidobro}, \citenamefont {Galiffi}, \citenamefont
  {Mejía-Salazar}, \citenamefont {Palermo}, \citenamefont {Jacob},\ and\
  \citenamefont {Maccaferri}}]{https://doi.org/10.48550/arxiv.2306.01314}%
  \BibitemOpen
  \bibfield  {author} {\bibinfo {author} {\bibfnamefont {M.}~\bibnamefont
  {Lobet}}, \bibinfo {author} {\bibfnamefont {N.}~\bibnamefont {Kinsey}},
  \bibinfo {author} {\bibfnamefont {I.}~\bibnamefont {Liberal}}, \bibinfo
  {author} {\bibfnamefont {H.}~\bibnamefont {Caglayan}}, \bibinfo {author}
  {\bibfnamefont {P.}~\bibnamefont {Arroyo-Huidobro}}, \bibinfo {author}
  {\bibfnamefont {E.}~\bibnamefont {Galiffi}}, \bibinfo {author} {\bibfnamefont
  {J.~R.}\ \bibnamefont {Mejía-Salazar}}, \bibinfo {author} {\bibfnamefont
  {G.}~\bibnamefont {Palermo}}, \bibinfo {author} {\bibfnamefont
  {Z.}~\bibnamefont {Jacob}}, \ and\ \bibinfo {author} {\bibfnamefont
  {N.}~\bibnamefont {Maccaferri}},\ }\href {\doibase 10.48550/ARXIV.2306.01314}
  {\enquote {\bibinfo {title} {New horizons in near-zero refractive index
  photonics and hyperbolic metamaterials},}\ } (\bibinfo {year}
  {2023})\BibitemShut {NoStop}%
\bibitem [{\citenamefont {Liu}\ \emph {et~al.}(2021)\citenamefont {Liu},
  \citenamefont {Zha}, \citenamefont {Shafi}, \citenamefont {Li}, \citenamefont
  {Yang}, \citenamefont {Xu}, \citenamefont {Liu},\ and\ \citenamefont
  {Jiang}}]{Liu2021}%
  \BibitemOpen
  \bibfield  {author} {\bibinfo {author} {\bibfnamefont {R.}~\bibnamefont
  {Liu}}, \bibinfo {author} {\bibfnamefont {Z.}~\bibnamefont {Zha}}, \bibinfo
  {author} {\bibfnamefont {M.}~\bibnamefont {Shafi}}, \bibinfo {author}
  {\bibfnamefont {C.}~\bibnamefont {Li}}, \bibinfo {author} {\bibfnamefont
  {W.}~\bibnamefont {Yang}}, \bibinfo {author} {\bibfnamefont {S.}~\bibnamefont
  {Xu}}, \bibinfo {author} {\bibfnamefont {M.}~\bibnamefont {Liu}}, \ and\
  \bibinfo {author} {\bibfnamefont {S.}~\bibnamefont {Jiang}},\ }\href
  {\doibase 10.1515/nanoph-2021-0301} {\bibfield  {journal} {\bibinfo
  {journal} {Nanophotonics}\ }\textbf {\bibinfo {volume} {10}},\ \bibinfo
  {pages} {2949} (\bibinfo {year} {2021})}\BibitemShut {NoStop}%
\bibitem [{\citenamefont {Zhao}\ \emph {et~al.}(2021)\citenamefont {Zhao},
  \citenamefont {Hubarevich}, \citenamefont {Iarossi}, \citenamefont {Borzda},
  \citenamefont {Tantussi}, \citenamefont {Huang},\ and\ \citenamefont
  {Angelis}}]{Zhao2021}%
  \BibitemOpen
  \bibfield  {author} {\bibinfo {author} {\bibfnamefont {Y.}~\bibnamefont
  {Zhao}}, \bibinfo {author} {\bibfnamefont {A.}~\bibnamefont {Hubarevich}},
  \bibinfo {author} {\bibfnamefont {M.}~\bibnamefont {Iarossi}}, \bibinfo
  {author} {\bibfnamefont {T.}~\bibnamefont {Borzda}}, \bibinfo {author}
  {\bibfnamefont {F.}~\bibnamefont {Tantussi}}, \bibinfo {author}
  {\bibfnamefont {J.-A.}\ \bibnamefont {Huang}}, \ and\ \bibinfo {author}
  {\bibfnamefont {F.~D.}\ \bibnamefont {Angelis}},\ }\href {\doibase
  10.1002/adom.202100888} {\bibfield  {journal} {\bibinfo  {journal} {Advanced
  Optical Materials}\ }\textbf {\bibinfo {volume} {9}},\ \bibinfo {pages}
  {2100888} (\bibinfo {year} {2021})}\BibitemShut {NoStop}%
\bibitem [{\citenamefont {Shafi}\ \emph {et~al.}(2021)\citenamefont {Shafi},
  \citenamefont {Liu}, \citenamefont {Zha}, \citenamefont {Li}, \citenamefont
  {Du}, \citenamefont {Wali}, \citenamefont {Jiang}, \citenamefont {Man},\ and\
  \citenamefont {Liu}}]{Shafi2021}%
  \BibitemOpen
  \bibfield  {author} {\bibinfo {author} {\bibfnamefont {M.}~\bibnamefont
  {Shafi}}, \bibinfo {author} {\bibfnamefont {R.}~\bibnamefont {Liu}}, \bibinfo
  {author} {\bibfnamefont {Z.}~\bibnamefont {Zha}}, \bibinfo {author}
  {\bibfnamefont {C.}~\bibnamefont {Li}}, \bibinfo {author} {\bibfnamefont
  {X.}~\bibnamefont {Du}}, \bibinfo {author} {\bibfnamefont {S.}~\bibnamefont
  {Wali}}, \bibinfo {author} {\bibfnamefont {S.}~\bibnamefont {Jiang}},
  \bibinfo {author} {\bibfnamefont {B.}~\bibnamefont {Man}}, \ and\ \bibinfo
  {author} {\bibfnamefont {M.}~\bibnamefont {Liu}},\ }\href {\doibase
  10.1016/j.apsusc.2021.149729} {\bibfield  {journal} {\bibinfo  {journal}
  {Applied Surface Science}\ }\textbf {\bibinfo {volume} {555}},\ \bibinfo
  {pages} {149729} (\bibinfo {year} {2021})}\BibitemShut {NoStop}%
\bibitem [{\citenamefont {Wang}\ \emph {et~al.}(2021)\citenamefont {Wang},
  \citenamefont {Huo}, \citenamefont {Ning}, \citenamefont {Liu}, \citenamefont
  {Zha}, \citenamefont {Shafi}, \citenamefont {Li}, \citenamefont {Li},
  \citenamefont {Xing}, \citenamefont {Zhang}, \citenamefont {Xu},
  \citenamefont {Li},\ and\ \citenamefont {Jiang}}]{Wang2021}%
  \BibitemOpen
  \bibfield  {author} {\bibinfo {author} {\bibfnamefont {Z.}~\bibnamefont
  {Wang}}, \bibinfo {author} {\bibfnamefont {Y.}~\bibnamefont {Huo}}, \bibinfo
  {author} {\bibfnamefont {T.}~\bibnamefont {Ning}}, \bibinfo {author}
  {\bibfnamefont {R.}~\bibnamefont {Liu}}, \bibinfo {author} {\bibfnamefont
  {Z.}~\bibnamefont {Zha}}, \bibinfo {author} {\bibfnamefont {M.}~\bibnamefont
  {Shafi}}, \bibinfo {author} {\bibfnamefont {C.}~\bibnamefont {Li}}, \bibinfo
  {author} {\bibfnamefont {S.}~\bibnamefont {Li}}, \bibinfo {author}
  {\bibfnamefont {K.}~\bibnamefont {Xing}}, \bibinfo {author} {\bibfnamefont
  {R.}~\bibnamefont {Zhang}}, \bibinfo {author} {\bibfnamefont
  {S.}~\bibnamefont {Xu}}, \bibinfo {author} {\bibfnamefont {Z.}~\bibnamefont
  {Li}}, \ and\ \bibinfo {author} {\bibfnamefont {S.}~\bibnamefont {Jiang}},\
  }\href {\doibase 10.3390/nano11030587} {\bibfield  {journal} {\bibinfo
  {journal} {Nanomaterials}\ }\textbf {\bibinfo {volume} {11}},\ \bibinfo
  {pages} {587} (\bibinfo {year} {2021})}\BibitemShut {NoStop}%
\bibitem [{\citenamefont {Wang}\ \emph {et~al.}(2022)\citenamefont {Wang},
  \citenamefont {Krasavin}, \citenamefont {Liu}, \citenamefont {Jiang},
  \citenamefont {Li}, \citenamefont {Guo}, \citenamefont {Tong},\ and\
  \citenamefont {Zayats}}]{Wang2022}%
  \BibitemOpen
  \bibfield  {author} {\bibinfo {author} {\bibfnamefont {P.}~\bibnamefont
  {Wang}}, \bibinfo {author} {\bibfnamefont {A.~V.}\ \bibnamefont {Krasavin}},
  \bibinfo {author} {\bibfnamefont {L.}~\bibnamefont {Liu}}, \bibinfo {author}
  {\bibfnamefont {Y.}~\bibnamefont {Jiang}}, \bibinfo {author} {\bibfnamefont
  {Z.}~\bibnamefont {Li}}, \bibinfo {author} {\bibfnamefont {X.}~\bibnamefont
  {Guo}}, \bibinfo {author} {\bibfnamefont {L.}~\bibnamefont {Tong}}, \ and\
  \bibinfo {author} {\bibfnamefont {A.~V.}\ \bibnamefont {Zayats}},\ }\href
  {\doibase 10.1021/acs.chemrev.2c00333} {\bibfield  {journal} {\bibinfo
  {journal} {Chemical Reviews}\ }\textbf {\bibinfo {volume} {122}},\ \bibinfo
  {pages} {15031} (\bibinfo {year} {2022})}\BibitemShut {NoStop}%
\bibitem [{\citenamefont {Liu}\ \emph {et~al.}(2023)\citenamefont {Liu},
  \citenamefont {Li}, \citenamefont {Du}, \citenamefont {Gao}, \citenamefont
  {Feng}, \citenamefont {Shafi}, \citenamefont {Jiang},\ and\ \citenamefont
  {Yue}}]{Liu2023}%
  \BibitemOpen
  \bibfield  {author} {\bibinfo {author} {\bibfnamefont {C.}~\bibnamefont
  {Liu}}, \bibinfo {author} {\bibfnamefont {L.}~\bibnamefont {Li}}, \bibinfo
  {author} {\bibfnamefont {X.}~\bibnamefont {Du}}, \bibinfo {author}
  {\bibfnamefont {J.}~\bibnamefont {Gao}}, \bibinfo {author} {\bibfnamefont
  {J.}~\bibnamefont {Feng}}, \bibinfo {author} {\bibfnamefont {M.}~\bibnamefont
  {Shafi}}, \bibinfo {author} {\bibfnamefont {S.}~\bibnamefont {Jiang}}, \ and\
  \bibinfo {author} {\bibfnamefont {W.}~\bibnamefont {Yue}},\ }\href {\doibase
  10.1016/j.optlastec.2023.109394} {\bibfield  {journal} {\bibinfo  {journal}
  {Optics {\&} Laser Technology}\ }\textbf {\bibinfo {volume} {163}},\ \bibinfo
  {pages} {109394} (\bibinfo {year} {2023})}\BibitemShut {NoStop}%
\bibitem [{\citenamefont {Peragut}\ \emph {et~al.}(2017)\citenamefont
  {Peragut}, \citenamefont {Cerruti}, \citenamefont {Baranov}, \citenamefont
  {Hugonin}, \citenamefont {Taliercio}, \citenamefont {Wilde},\ and\
  \citenamefont {Greffet}}]{Peragut2017}%
  \BibitemOpen
  \bibfield  {author} {\bibinfo {author} {\bibfnamefont {F.}~\bibnamefont
  {Peragut}}, \bibinfo {author} {\bibfnamefont {L.}~\bibnamefont {Cerruti}},
  \bibinfo {author} {\bibfnamefont {A.}~\bibnamefont {Baranov}}, \bibinfo
  {author} {\bibfnamefont {J.~P.}\ \bibnamefont {Hugonin}}, \bibinfo {author}
  {\bibfnamefont {T.}~\bibnamefont {Taliercio}}, \bibinfo {author}
  {\bibfnamefont {Y.~D.}\ \bibnamefont {Wilde}}, \ and\ \bibinfo {author}
  {\bibfnamefont {J.~J.}\ \bibnamefont {Greffet}},\ }\href {\doibase
  10.1364/optica.4.001409} {\bibfield  {journal} {\bibinfo  {journal} {Optica}\
  }\textbf {\bibinfo {volume} {4}},\ \bibinfo {pages} {1409} (\bibinfo {year}
  {2017})}\BibitemShut {NoStop}%
\bibitem [{\citenamefont {Lai}\ \emph {et~al.}(2017)\citenamefont {Lai},
  \citenamefont {Wang}, \citenamefont {Ling}, \citenamefont {Wang},
  \citenamefont {kai Chiu}, \citenamefont {Chau}, \citenamefont {Huang},\ and\
  \citenamefont {Chiang}}]{Lai2017}%
  \BibitemOpen
  \bibfield  {author} {\bibinfo {author} {\bibfnamefont {C.-H.}\ \bibnamefont
  {Lai}}, \bibinfo {author} {\bibfnamefont {G.-A.}\ \bibnamefont {Wang}},
  \bibinfo {author} {\bibfnamefont {T.-K.}\ \bibnamefont {Ling}}, \bibinfo
  {author} {\bibfnamefont {T.-J.}\ \bibnamefont {Wang}}, \bibinfo {author}
  {\bibfnamefont {P.}~\bibnamefont {kai Chiu}}, \bibinfo {author}
  {\bibfnamefont {Y.-F.~C.}\ \bibnamefont {Chau}}, \bibinfo {author}
  {\bibfnamefont {C.-C.}\ \bibnamefont {Huang}}, \ and\ \bibinfo {author}
  {\bibfnamefont {H.-P.}\ \bibnamefont {Chiang}},\ }\href {\doibase
  10.1038/s41598-017-05939-0} {\bibfield  {journal} {\bibinfo  {journal}
  {Scientific Reports}\ }\textbf {\bibinfo {volume} {7}} (\bibinfo {year}
  {2017}),\ 10.1038/s41598-017-05939-0}\BibitemShut {NoStop}%
\bibitem [{\citenamefont {Starko-Bowes}\ \emph {et~al.}(2015)\citenamefont
  {Starko-Bowes}, \citenamefont {Atkinson}, \citenamefont {Newman},
  \citenamefont {Hu}, \citenamefont {Kallos}, \citenamefont {Palikaras},
  \citenamefont {Fedosejevs}, \citenamefont {Pramanik},\ and\ \citenamefont
  {Jacob}}]{StarkoBowes2015}%
  \BibitemOpen
  \bibfield  {author} {\bibinfo {author} {\bibfnamefont {R.}~\bibnamefont
  {Starko-Bowes}}, \bibinfo {author} {\bibfnamefont {J.}~\bibnamefont
  {Atkinson}}, \bibinfo {author} {\bibfnamefont {W.}~\bibnamefont {Newman}},
  \bibinfo {author} {\bibfnamefont {H.}~\bibnamefont {Hu}}, \bibinfo {author}
  {\bibfnamefont {T.}~\bibnamefont {Kallos}}, \bibinfo {author} {\bibfnamefont
  {G.}~\bibnamefont {Palikaras}}, \bibinfo {author} {\bibfnamefont
  {R.}~\bibnamefont {Fedosejevs}}, \bibinfo {author} {\bibfnamefont
  {S.}~\bibnamefont {Pramanik}}, \ and\ \bibinfo {author} {\bibfnamefont
  {Z.}~\bibnamefont {Jacob}},\ }\href {\doibase 10.1364/josab.32.002074}
  {\bibfield  {journal} {\bibinfo  {journal} {Journal of the Optical Society of
  America B}\ }\textbf {\bibinfo {volume} {32}},\ \bibinfo {pages} {2074}
  (\bibinfo {year} {2015})}\BibitemShut {NoStop}%
\bibitem [{\citenamefont {Khan}\ \emph {et~al.}(2022)\citenamefont {Khan},
  \citenamefont {Khan}, \citenamefont {Xie}, \citenamefont {Abbas},
  \citenamefont {Rauf}, \citenamefont {Mehmood}, \citenamefont {Runowski},
  \citenamefont {Agathopoulos},\ and\ \citenamefont {Zhu}}]{Khan2022}%
  \BibitemOpen
  \bibfield  {author} {\bibinfo {author} {\bibfnamefont {S.~A.}\ \bibnamefont
  {Khan}}, \bibinfo {author} {\bibfnamefont {N.~Z.}\ \bibnamefont {Khan}},
  \bibinfo {author} {\bibfnamefont {Y.}~\bibnamefont {Xie}}, \bibinfo {author}
  {\bibfnamefont {M.~T.}\ \bibnamefont {Abbas}}, \bibinfo {author}
  {\bibfnamefont {M.}~\bibnamefont {Rauf}}, \bibinfo {author} {\bibfnamefont
  {I.}~\bibnamefont {Mehmood}}, \bibinfo {author} {\bibfnamefont
  {M.}~\bibnamefont {Runowski}}, \bibinfo {author} {\bibfnamefont
  {S.}~\bibnamefont {Agathopoulos}}, \ and\ \bibinfo {author} {\bibfnamefont
  {J.}~\bibnamefont {Zhu}},\ }\href {\doibase 10.1002/adom.202200500}
  {\bibfield  {journal} {\bibinfo  {journal} {Advanced Optical Materials}\
  }\textbf {\bibinfo {volume} {10}},\ \bibinfo {pages} {2200500} (\bibinfo
  {year} {2022})}\BibitemShut {NoStop}%
\bibitem [{\citenamefont {Ahmadivand}\ and\ \citenamefont
  {Gerislioglu}(2021)}]{Ahmadivand2021}%
  \BibitemOpen
  \bibfield  {author} {\bibinfo {author} {\bibfnamefont {A.}~\bibnamefont
  {Ahmadivand}}\ and\ \bibinfo {author} {\bibfnamefont {B.}~\bibnamefont
  {Gerislioglu}},\ }\href {\doibase 10.1002/lpor.202100328} {\bibfield
  {journal} {\bibinfo  {journal} {Laser {\&} Photonics Reviews}\ }\textbf
  {\bibinfo {volume} {16}},\ \bibinfo {pages} {2100328} (\bibinfo {year}
  {2021})}\BibitemShut {NoStop}%
\bibitem [{\citenamefont {Sahoo}\ \emph {et~al.}(2023)\citenamefont {Sahoo},
  \citenamefont {VS},\ and\ \citenamefont {Kumar}}]{Sahoo2023}%
  \BibitemOpen
  \bibfield  {author} {\bibinfo {author} {\bibfnamefont {M.~K.}\ \bibnamefont
  {Sahoo}}, \bibinfo {author} {\bibfnamefont {A.~A.}\ \bibnamefont {VS}}, \
  and\ \bibinfo {author} {\bibfnamefont {A.}~\bibnamefont {Kumar}},\ }\href
  {\doibase 10.1088/1361-6528/acb948} {\bibfield  {journal} {\bibinfo
  {journal} {Nanotechnology}\ }\textbf {\bibinfo {volume} {34}},\ \bibinfo
  {pages} {195301} (\bibinfo {year} {2023})}\BibitemShut {NoStop}%
\bibitem [{\citenamefont {Vasilantonakis}\ \emph {et~al.}(2015)\citenamefont
  {Vasilantonakis}, \citenamefont {Wurtz}, \citenamefont {Podolskiy},\ and\
  \citenamefont {Zayats}}]{Vasilantonakis2015}%
  \BibitemOpen
  \bibfield  {author} {\bibinfo {author} {\bibfnamefont {N.}~\bibnamefont
  {Vasilantonakis}}, \bibinfo {author} {\bibfnamefont {G.~A.}\ \bibnamefont
  {Wurtz}}, \bibinfo {author} {\bibfnamefont {V.~A.}\ \bibnamefont
  {Podolskiy}}, \ and\ \bibinfo {author} {\bibfnamefont {A.~V.}\ \bibnamefont
  {Zayats}},\ }\href {\doibase 10.1364/oe.23.014329} {\bibfield  {journal}
  {\bibinfo  {journal} {Optics Express}\ }\textbf {\bibinfo {volume} {23}},\
  \bibinfo {pages} {14329} (\bibinfo {year} {2015})}\BibitemShut {NoStop}%
\end{thebibliography}%

\noindent 

\section*{Acknowledgements}
A.K. acknowledges funding support from the Department of Science and Technology via the grants: SB/S2/RJN-110/2017, ECR/2018/001485, and DST/NM/NS-2018/49. We acknowledge the Centre of Excellence in Nanoelectronics (CEN) at IIT Bombay for providing fabrication and characterization facilities. M.K.S. acknowledges the Institute Postdoctoral fellowship for financial support. A.A. acknowledges UGC for his junior research fellowship to support his Ph.D. N.R.S. acknowledges CSIR fellowship to support his Ph.D.

\end{document}